\documentclass[conference]{IEEEtran}
\IEEEoverridecommandlockouts
\ifCLASSINFOpdf
\else
\fi
\usepackage{amsmath}

\usepackage[ruled, lined, linesnumbered, commentsnumbered, longend]{algorithm2e}
\usepackage{xcolor}
\usepackage{graphicx}

\begin{document}
%
\title{Red Teaming Methodology for Design Obfuscation}
%
%
%

\author{\IEEEauthorblockN{Yuntao Liu, Abir Akib, Zelin Lu, Qian Xu, Ankur Srivastava, Gang Qu}
\IEEEauthorblockA{Institute for Systems Research\\
University of Maryland, College Park\\
College Park, MD 20742\\
Email: \{ankurs,gangqu\}@umd.edu}
\and
\IEEEauthorblockN{David Kehlet, Nij Dorairaj}
\IEEEauthorblockA{Intel Corporation\\
San Jose, CA 95134\\
Email: \{david.kehlet, nij.dorairaj\}@intel.com}
%
%
%
%
\thanks{DISTRIBUTION STATEMENT A. Approved for public release: distribution is unlimited.}
}

\maketitle

\begin{abstract}
The main goal of design obfuscation schemes is to protect sensitive design details from untrusted parties in the VLSI supply chain, including but not limited to off-shore foundries and untrusted end users. In this work, we provide a systematic red teaming approach to evaluate the security of design obfuscation approaches. Specifically, we propose security metrics and evaluation methodology for the scenarios where the adversary does not have access to a working chip. A case study on the RIPPER tool developed by the University of Florida \cite{dasgupta2022ripper} indicates that more information is leaked about the structure of the original design than commonly considered.
\end{abstract}


\section{Introduction}


Due to the increasing cost of maintaining IC foundries with advanced technology nodes, many chip designers have become fabless and outsource their fabrication to off-shore foundries. However, such foundries are not under the designer's control which puts the security of the IC supply chain at risk. Similarly, malicious end users or other adversaries may also try to steal the design secrets in fabricated ICs.
Many design-for-trust techniques have been studied as countermeasures among which design obfuscation has been the most widely studied \cite{chakraborty2019keynote}. 
Most design obfuscation techniques require adding a secret key input to the obfuscated design. The functionality of the circuit is correct only if the key is correct. The correct key is kept only by the designer and not known to the untrusted parties, hence preventing the actual functionality of the design from being leaked.

The evolution of design obfuscation techniques have been following a ping-pong approach where point defenses are followed by point attacks leading to a never ending flurry of activities. 
Due to this trend, the main goal of many recently proposed obfuscation techniques has been counteracting the latest attack techniques. A comprehensive evaluation methodology on design obfuscation is yet to be developed. 

Innovations in design obfuscation technology require a systematic and independent red teaming approach. 
Specifically, a basic framework that captures different ways in which attacks can be launched should be created. The framework should also be extendable for newly identified attack surfaces. 

In our work, we attempt to standardize the way obfuscation technologies are evaluated. 
The security level obfuscation is captured by how much uncertainty in the functionality the obfuscation can cause. This can be quantified by \textbf{the number of possible functionalities} that the obfuscated design can represent by having different correct keys under each attack scenario.
Some attacks require knowledge beyond the obfuscated netlist, such as prior knowledge of the design or a working chip. Therefore, we identify 3 attack scenarios under which the security level of the obfuscation technique should be evaluated:
\begin{enumerate}
    \item The adversary only has the obfuscated design netlist.
    \item The adversary has some prior knowledge about the design in addition to the obfuscated netlist.
    \item The adversary also has a working chip from which the adversary can observe correct input-output pairs.
\end{enumerate}



In this paper, we focus on the first two scenarios, \textit{i. e.,} the adversary \textit{does not} have access to a working system. While the case with working system access is also important, the case without a working system is much less investigated is much less investigated in prior art. The contribution of this paper is as follows:
\begin{itemize}
    \item For each scenario, we identify possible sources information leakage and attack techniques. 
    \item Based on sources of leakage, we develop systematic red teaming approach to evaluate the security of design obfuscation techniques.
    \item As a case study, we evaluate the RIPPER tool developed by the University of Florida \cite{dasgupta2022ripper} using our approach. Experiment results show that there exist information leakage in both scenarios, and the leakage is much more severe if prior knowledge is available for the obfuscated design.
\end{itemize}

The main objective of this paper is to help the design obfuscation community standardize how it approaches formal evaluations of design obfuscation solutions.

\section{Related Work}
\label{sec:related_work}
The earliest design obfuscation techniques aimed at inserting substantial error into the circuit with a wrong key \cite{roy2008epic, rajendran2012security, rajendran2015fault}. 
Since then, various attacks on obfuscation techniques that need or do not need a working chip have been exploited in the literature. 
The Boolean satisfiability (SAT) based attack \cite{subramanyan2015evaluating} is the first attack that broke all the obfuscation techniques that preceded it and it needs a working chip. A set of SAT-resistant obfuscation approaches, such as SARLock~\cite{yasin2016sarlock} and Anti-SAT~\cite{xie2018anti} forced the SAT attacks to undergo exponential complexity. These are also followed by new attacks and newer defenses and the cat-and-mouse game is still ongoing.

Attacks that do not require a working system have also gained traction recently. Desynthesis attack \cite{https://doi.org/10.48550/arxiv.1703.10187} exploits biases in obfuscation tool to determine if a specific key yields a functionality which when is re-obfuscated using the same tool gives a similar locked netlist. SAIL- structural analysis using machine learning \cite{SAIL} is another structural attack developed based on the fact that obfuscation tools introduce sparse and local changes which are very deterministic. 
SWEEP attack \cite{SWEEP} trains a machine learning model to find correlations between the correct key value and features extracted from synthesis reports.
Although these attacks do not need a working chip, they mostly target rudimentary obfuscation schemes based on XOR/XNOR or multiplexer insertion which are not secure against SAT-based attacks and still leave the netlist mostly exposed.

\section{Proposed Red Teaming Approach}
In this work, we propose a systematic red teaming approaches for logic obfuscation. To do this, we develop specific security evaluation frameworks for the first two scenarios.

\subsection{Threat Model and Security Metric}
Each scenario defines a different threat model with varied accesses to resource. As the goal of the adversary is usually to find out the actual functionality of the obfuscated design, we use \textbf{the total number of possible functionalities} in the obfuscated netlist as the security metric.





\subsection{Security Evaluation under Scenario 1}
Scenario 1 is the case for an adversary who has only the locked reverse engineered netlist without any access to oracle or prior knowledge about the functionality of the circuit. 
The adversary can further reduce the functionality search space using the following information:
\begin{enumerate}
    \item The unaltered portion of the design, including but not limited to circuit topology and partial functionality. 
    \item Biases of the obfuscation algorithm. Such bias will limit the key search space and hence that of functionality.
\end{enumerate}
As discussed in Sec. \ref{sec:related_work}, existing attacks have shown that these sources of information leakage reveal most of the key bits of rudimentary obfuscation techniques. 
We will analyze the security level of RIPPER under this scenario.
Each LUT correspond to one or more gates in the original netlist. \textit{This leaks the original topology of the design.}
This also means that the function represented by the LUT must contain all its input bits, \textit{i. e.,} there is no don't-care input for any LUT. For example, let us examine a 2-input LUT with inputs $a_1,a_0$ and configuration bits $m_3,\ldots,m_0$. Among the 16 functionalities that this LUT can represent, 6 functionalities have don't-care inputs (namely constants 0 and 1, $a_1$, $\bar{a_1}$, $a_0$, $\bar{a_0}$), which reduces the number of legitimate functions to 10. 
Similarly, a 3-input LUT and a 4-input LUT can represent 218 and 64595 functionalities without don't-care inputs, respectively.

Let us further consider a single-output, tree-shaped logic cone composed of only LUTs without bitstream compaction. Let $n_i$ be the number of $i$-input LUTs in the cone. Currently in the RIPPER tool, a LUT can have 2 to 4 inputs so $i$ can be 2, 3, or 4. Such a logic cone has $\sum_{i=2}^4(i-1)a_i+1$ input bits and $\sum_{i=2}^42^ia_i$ configuration bits. Let us use $F_I$, $F_K$, and $F_R$ to represent the maximum number of functionalities that can be implemented by the inputs, the configuration bits (keys), and under RIPPER, respectively. We have:
\begin{align}
    F_I =&\ 2^{2^{\sum_{i=2}^4(i-1)a_i+1}},\\ 
    F_K =&\ 2^{\sum_{i=2}^42^ia_i},\\
    F_R =&\ \frac{\prod_{i=2}^4b_i^{a_i}}{2^{\sum_{i=2}^4a_i-1}}
    \label{eq:num_funcs}
\end{align}
where $b_i$ stand for the number of eligible functionalities that can be implemented by an $i$-input LUT, \textit{i.e. } $b_2=10, b_3=218, b_4=64595$. $F_I$ and $F_K$ are relatively straightforward. $F_R$ is the case because each legitimate functionality under RIPPER's constraints that there cannot be don't-care inputs in any LUT. 
The numerator term indicates the number of valid bitstreams under this constraints. The denominator term quantifies how many equivalent bitstreams can implement each valid functionality. Due to De Morgan's Law, given a valid bitstream that implement any functionality, each LUT that is not the output of the cone can have all its configuration bits flipped, and there is always a way to adjust the configuration bits in its following LUT to accomodate for the flipping and maintain the same functionality. Hence, for each valid functionality, there are always $2^{\sum_{i=2}^4a_i-1}$ bitstreams that can implement it.

\begin{figure}[htb]
    \centering
  \vspace{-1mm}    \includegraphics[width=.48\textwidth]{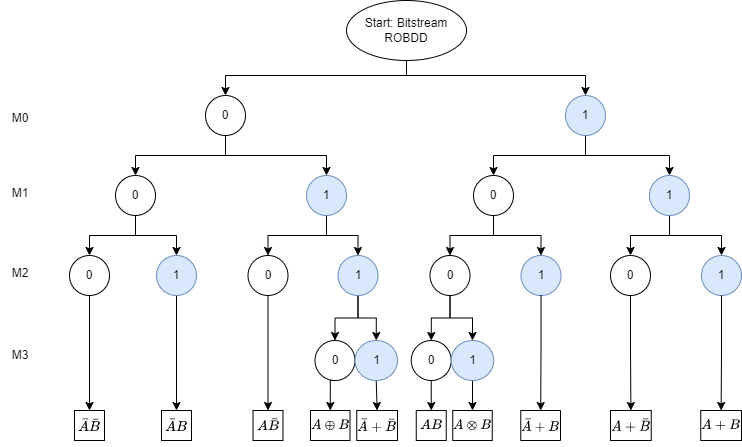}
    \caption{ROBDD for a 2-input LUT under no don't-care input requirement}
    \label{fig:bdd}
\end{figure}

In Table \ref{tab:num_funcs}, we show 3 types of redacted netlists that have 16 configuration bits in total: 4 2-input LUTs, 2 2-input LUTs and 1 3-input LUT, and 1 4-input LUT. The first two have 5 input bits and the last one has 4 input bits. Although the last one has a lower $F_I$ due to one fewer input, it has the largest $F_R$. This is because a 4-input LUT can implement any 4-input Boolean function. Except a small portion of the functionalities that have don't-care inputs, most of them remain valid in RIPPER. This indicates that, if the number of configuration bits is limited, using larger LUTs can increase the number of functionalities and improve security.
\vspace{-1mm}
\begin{table}[htb]
    \centering
    \caption{Comparison of the Number of Functionalities}
    \begin{tabular}{|c|c|c|c|c|c|}
        \hline
        $a_2$ & $a_3$ & $a_4$ & $F_I$ & $F_K$ & $F_R$ \\ \hline
        $4$ & 0 & 0 & $4.29\times 10^9$ & 65,536 & 1,250 \\ \hline
        $2$ & 1 & 0 & $4.29\times 10^9$ & 65,536 & 5,450 \\ \hline
        $0$ & 0 & 1 & 65,536 & 65,536 & 64,595 \\ \hline
    \end{tabular}
    \label{tab:num_funcs}
\end{table}
\vspace{-1mm}
It is noteworthy that the $F_R$ expression as defined in Eq. \ref{eq:num_funcs} is suitable only for redacted netlists that are tree-shaped and without bitstream compaction (\textit{i. e. } sharing of configuration bits across different LUT input pins. For a generic netlist, it is hard to express $F_R$ with a concise analytical formula due to the variations in design topology and the way keys are shared among different key ports (such as bitstream compaction in RIPPER). Hence, we propose an algorithm that can identify each unique functionality that can be implemented by an obfuscated design. 
Specifically, we use a reduced-order binary decision diagram (ROBDD) to search for legitimate bitstreams under RIPPER's constraints. ROBDD is a binary tree diagram commonly used construct to express a Boolean function's functionality. In a conventional ROBDD, each layer of the diagram indicate the value of one input of the netlist, and the last layer indicate the output value of the circuit. 
In our ROBDD, each layer indicate the value of one configuration bit, and the last layer indicates the functionality that the redacted netlist implements with the bitstream. In Figure \ref{fig:bdd}, we illustrate an ROBDD for a 2-input LUT.

\vspace{-1mm}

\begin{figure}[htb]
    \centering
    \includegraphics[width=.48\textwidth]{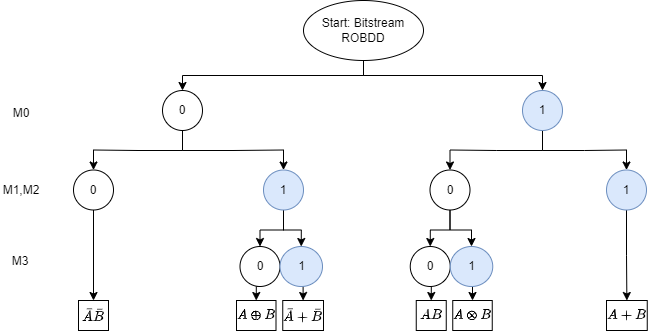}
    \caption{ROBDD for a 2-input LUT with two configuration bets shared}
    \label{fig:compaction_bdd}
\end{figure}

The ROBDD representation offers great flexibility in implementing the constraints on the bitstreams. For example, we can exclude the branches that does not conform with the no-don't-care constraint. For bitstreams, the compacted layers can be combined into one layer as their values must be the same. In Figure \ref{fig:compaction_bdd}, we demonstrate the ROBDD for a 2-input LUT where two configuration bits are compacted. By condensing the compacted layers, we can obtain the new ROBDD for the compacted bitstream.

In Sec. \ref{ssec:exp_1}, we provide experimental data on the number of possible functionalities of netlists redacted by RIPPER.

\subsection{Security Evaluation under Scenario 2}

In Scenario 2, the adversary has some prior knowledge about the design. Few previous study has considered this scenario. Real-world adversaries are likely experienced engineers and/or institutions who are knowledgeable about chip designs. Due the inherent subjectivity of the prior knowledge, the prior knowledge of the adversary is very difficult to characterize mathematically. 
As the first attempt in this direction, we use a set of benchmark circuits to represent the knowledge of an adversary. 
Specifically, we assume that the adversary only knows the library designs' RTL and do not know which synthesis tool is used to obtain the netlist under attack. 
The library may or may not have the original design and may have an older version of the design. In the future, machine learning based approaches can be developed to obtain enriched design databases from a set of known designs.

We attempt to match the obfuscated netlists with known designs by identifying structural similarity between them.
Cheng \textit{et. al.} have proposed similarity-based circuit partitioning methods \cite{Partitioning} which has the capabilities of extracting similar subcircuits existing in two different netlists. Some commercial verification tools like Cadence Conformal Logic Equivalence Check (LEC) give more detailed report on functional similarities and dissimilarities between two netlist including where the dissimilarities may occur. These tools can be exploited to find which netlist from the library has highest structural and logical similarities with the netlist under consideration. That particular design from the library will be used for attacks to extract the keys.




\section{Experiments}

We evaluate our approach on the RIPPER tool \cite{dasgupta2022ripper}. RIPPER replaces each logic gate with a look-up table (LUT). Since the original structure of the netlist is preserved, information about the original design is leaked. Such leakage has different ramifications under each scenario.

\subsection{Scenario 1} \label{ssec:exp_1}

\begin{figure}[htb]
    \centering
    \includegraphics[width=.45\textwidth]{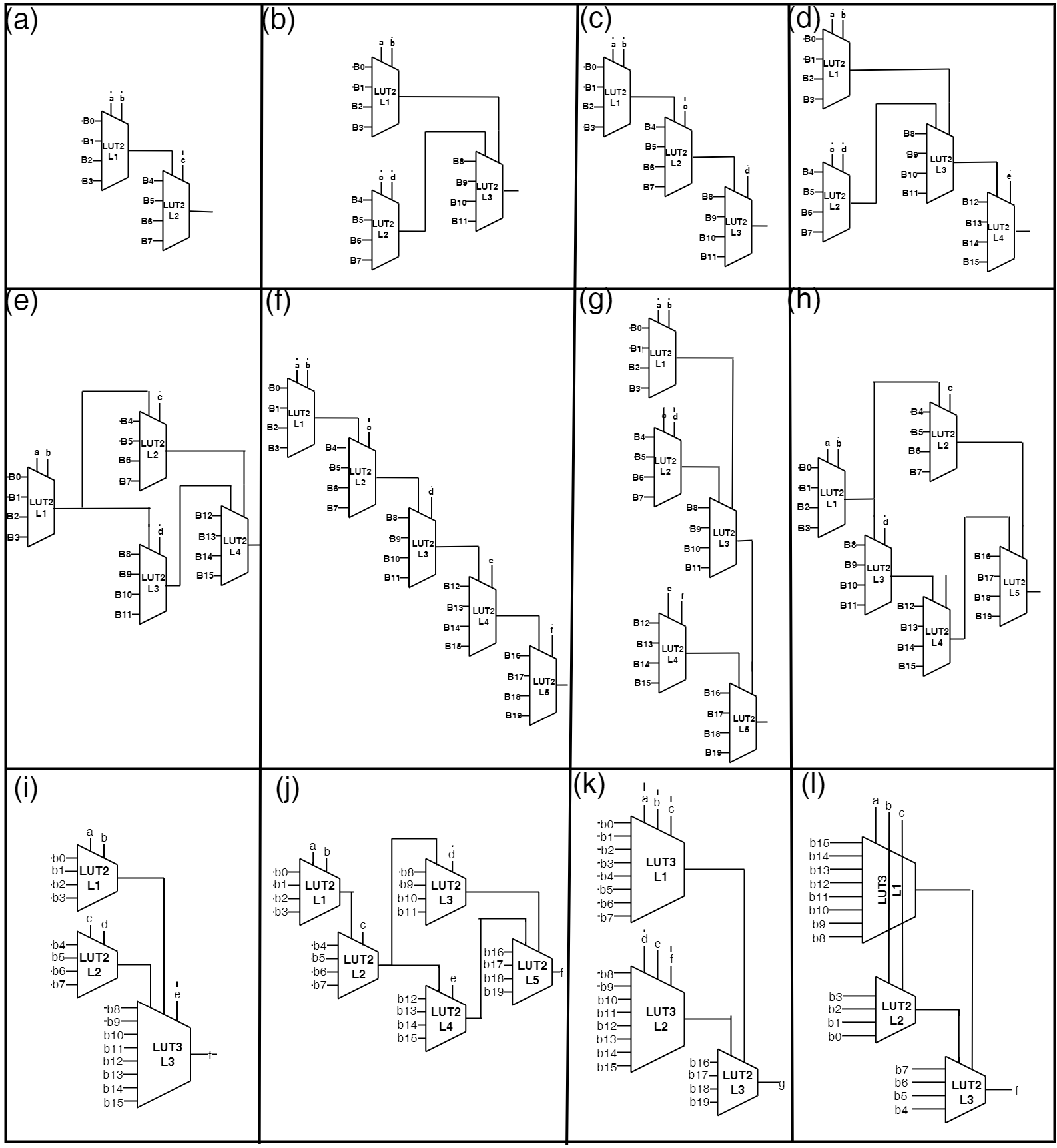}
    \vspace{-1mm}
    \caption{Different topologies with implementable unique functionality of (a) 50, (b) 250, (c) 250, (d) 1250, (e) 604, (f) 6250, (g) 6250, (h) 4124, (i) 5450, (j) 3004, (k) 118810, and (l) 256}
    \label{circuits}
\end{figure}

For Scenario 1, we examine the number of functionalities that can be possibly implemented by the redacted netlist produced by RIPPER. Figure \ref{circuits} shows some the topologies of the logic cones we have considered. We enumerate the possible functionalitis and show the numbers in the caption of Figure \ref{circuits}. As we see, the numbers are also consistent with $F_R$ as defined in Eq. \ref{eq:num_funcs}.

\vspace{-1mm}
\vspace{-1mm}

\subsection{Scenario 2}

Cadence Conformal Logic Equivalence Check (LEC) is a formal verification tool which is used to determine logical equivalence between two netlists. The tool operates in two stages – (a) finding key point mapping between the two circuits and (b) checking logical equivalence between the fan in cones of the key points. \textit{The key points are flipflops, floating pins, and IO pins.} Presence of key gates will cause the two circuits to be logically non-equivalent, provided the key is not known. Stage 1, which is the mapping of key points is of special interest as it exploits the structural similarities between the two circuits. Two circuits having same functionality will have more similarities than those having different functionalities. We have experimented with SHA256, SHA512 and GPS benchmarks and the result is illustrated in the table in \ref{tab:results}.

\vspace{-1mm}
\begin{table}[hbt]
\centering
\caption{\label{tab:results}Mapping profile of locked netlist vs RTL netlist from library.}
\begin{tabular}{ |c|c|c|c| } 
 \hline
 RTL Design & \multicolumn{3}{c|}{Fully redacted design} \\ \cline{2-4}
 Library & SHA256 & SHA512 & GPS\\
 \hline
 SHA256 & \textbf{Mapped} & Unmapped & Unmapped \\ 
 \hline
 SHA512 & Unmapped & \textbf{Mapped} & Unmapped \\ 
 \hline
 GPS & Unmapped & Unmapped & \textbf{Mapped} \\ 
 \hline
\end{tabular}
\end{table}
\vspace{-1mm}

Design methodology and obfuscation tools are inconsistent among different designers which might cause a design to look different than what an adversary might have in their library. Some of these inconsistencies have been investigated thoroughly for SHA 256 benchmark and Cadence Conformal LEC tool has deemed itself useful under all the scenarios.

\begin{itemize}
    \item \textbf{Nomenclature of internal nodes and IO pins:} Mapping profile of Cadence Conformal LEC is independent of the nomenclature of the key points
    \item \textbf{Extra logic circuits and pins:} Though extra logic circuits and pins introduce some unmapped key points, all the key points internal to SHA256 are still mapped suggesting that SHA256 was the source of the netlist.
    \item \textbf{Changes in internal bus structure:} The changes in internal bus structure has no effect in the mapping profile.
    \item \textbf{Reduction of a part of circuit} The experiments have been performed on circuits obfuscated using RIPPER techniques and replacing logic gates with look-up tables have showed no impact on mapping profile. 
\end{itemize}
The results promote Cadence Conformal LEC to be extremely useful in shortlisting potential candidates from the adversary's library to be used for key-extraction attacks.


\section{Conclusion}
In this paper, we propose systematic red teaming methodology for design obfuscation techniques. We use the number of possible functionalities that can be implemented by the obfuscated netlist as the security metric. We put emphasis on the attack scenarios where the adversary does not have a working chip, as such scenarios are less studied in prior art. A case study is performed on the logic redaction based obfuscation technique called RIPPER. When the adversary has no prior knowledge of the design (Scenario 1), the number of possible functionalities is significantly smaller than the maximum number that can be implemented by the same set of inputs or keys (configuration bits). When original form of the obfuscated design exists in the adversary's prior knowledge, existing commercial tools can map the obfuscated design with its original design. These findings indicate that prior research in attacks on logic obfuscation failed to capture the structural leakage of redaction based techniques like RIPPER and underscores the importance of systematic red teaming effort for obfuscation.

\section*{Acknowledgment}
This work is supported by the National Security Technology Accelerator (NSTXL) Consortium contract number N00164-19-9-G007b.
\footnotetext{©UNIVERSITY OF MARYLAND and INTEL CORPORATION. Intel, the Intel logo, and other Intel marks are trademarks of Intel Corporation or its subsidiaries. *Other names and brands may be claimed as the property of others. LEGAL DISCLAIMER: Intel technologies’ features and benefits depend on system configuration and may require enabled hardware, software or service activation. No computer system can be absolutely secure. Results have been estimated based on internal Intel analysis and are provided for informational purposes only. Any difference in system hardware or software design or configuration may affect actual performance. This document contains information on products, services and/or processes in development. All
information provided here is subject to change without notice. Contact your Intel representative to obtain the latest forecast, schedule, specifications and roadmaps.}

\bibliographystyle{IEEEtran}
\bibliography{ref}

\end{document}